\documentclass[preprint,
12pt,
amsmath,amssymb,preprintnumbers]
{revtex4}
\usepackage{graphicx}
\usepackage{color}

\newcommand{\bmp}{\noindent\begin{minipage}{16cm}}
\newcommand{\emp}{\end{minipage}\vskip 7mm} 

\usepackage{bm}
\usepackage{bbm}

\newcommand{\beq}{\begin{equation}}
\newcommand{\eeq}{\end{equation}}
\newcommand{\bea}{\begin{eqnarray}}
\newcommand{\eea}{\end{eqnarray}}
\newcommand{\ba}{\begin{array}}
\newcommand{\ea}{\end{array}}
\newcommand{\bi}{\begin{itemize}}
\newcommand{\ei}{\end{itemize}}
\newcommand{\bn}{\begin{enumerate}}
\newcommand{\en}{\end{enumerate}}
\newcommand{\bc}{\begin{center}}
\newcommand{\ec}{\end{center}}

\newcommand{\gsim}{\lower.7ex\hbox{$\;\stackrel{\textstyle>}{\sim}\;$}}
\newcommand{\lsim}{\lower.7ex\hbox{$\;\stackrel{\textstyle<}{\sim}\;$}}

\definecolor{rossoCP3}{cmyk}{0,.88,.77,.40}

\begin{document}

\title{\Large  \color{rossoCP3}  
Hints of a Charge Asymmetry \\ in the Electron and Positron Cosmic-Ray Excesses\\ \vskip 1.cm} 

\author{Isabella {\sc  Masina}$^{\color{rossoCP3}{\clubsuit},{\heartsuit}}$}

\author{Francesco {\sc Sannino}$^{\color{rossoCP3}{\heartsuit}}$}

\affiliation{ \vskip .3cm $^{\color{rossoCP3}{\clubsuit}}${\mbox{Dip.~di Fisica e Scienze della Terra, 
Universit\`a di Ferrara and INFN, Ferrara, Italy }}}


\affiliation{    $^{\color{rossoCP3}{\heartsuit}}${\mbox{CP$^{ \bf 3}$-Origins and DIAS, 
University of Southern Denmark,  Odense, Denmark
 }}}



\begin{abstract} \vskip .5cm 
\baselineskip=15pt
By combining the recent data from AMS-02 with those from Fermi-LAT, 
we show the emergence of a charge asymmetry in the electron and positron cosmic-ray  excesses, 
slightly favoring the electron component. 
Astrophysical and dark matter inspired models introduced to explain the observed excesses 
can be classified according to their prediction for the charge asymmetry and its energy dependence.
Future data confirming the presence of a charge asymmetry, would imply
that an asymmetric production mechanism is at play.
 \\[.1cm]
{\footnotesize  \it Preprint: CP3-Origins-2013-008 DNRF90  \& DIAS-2013-8}
\end{abstract}

\maketitle

\baselineskip=16pt
\setcounter{page}{1}

\newpage


\section{Introduction}

Electron and positron fluxes in Cosmic Rays (CRs) have been measured by many experiments. 
For energies above about $10$ GeV the positron fraction displays a rising behavior.
The  total flux also displays some features around $100$ GeV. 
There is a common agreement on the fact that these data cannot 
be interpreted solely in terms of known astrophysical sources. An unknown source of electrons and positrons
CRs has to be introduced in order to account  for the excesses seen on the top of the
background fluxes, associated to known astrophysical sources. 
We refer to \cite{Fan:2010yq, Panov:2013fma} for recent reviews. 

In \cite{Frandsen:2010mr}  we suggested a strategy to investigate the possible charge asymmetry 
in the electron and positron CRs excesses; as a result, we showed that, at that time,
even large deviations from charge symmetry were experimentally viable. 
This kind of analysis deserves now to be updated in the light of the recent data collected by AMS-02 \cite{AMS02}
and by Fermi-LAT \cite{Ackermann:2010ij, Ackermann:2011rq}. This is precisely the goal of this paper.
Future experimental observations could even better constrain the amount of charge asymmetry.  

The amount of the charge asymmetry of the CR lepton excesses is crucial to understand the physical properties
of the unknown source.  Among the many candidates suggested, there are for instance: 
astrophysical sources, like supernovae or pulsars 
(see {\it e.g.}  \cite{Fan:2010yq,Linden:2013mqa, Delahaye:2010ji, Serpico:2011wg, Panov:2013fma, Caprioli:2009fv}),
which are expected to be charge symmetric;
dark matter (DM) annihilations and/or decays.
Contrary to the case of annihilating DM, 
decaying DM can lead to a charge asymmetry \cite{Chang:2011xn,Masina:2011hu,Carone:2011ur}
provided that both charge conjugation and lepton flavor are violated \cite{Masina:2011hu}.

The paper is organized as follows. In sec. \ref{data} we review the experimental data to be used in the analysis.
Section \ref{sumrule} introduces the notations and a useful relation \cite{Frandsen:2010mr}. After having discussed
background models in sec. \ref{background}, in sec. \ref{asymm} we investigate the experimental status of the charge asymmetry. Here we find a hint of an asymmetric excess, favouring the electron component over the positron one. We conclude in sec. \ref{conclusions}, where we offer the physical applications and interpretations
of our results.  

\section{Data on Electron and Positron CRs Fluxes}
\label{data}

We now briefly summarize the experimental data on electron and positron CRs fluxes, $\phi_{e^-}(E)$ and  $\phi_{e^+}(E)$,
where $E$ is the energy of the detected $e^{\pm}$.
 
In 2009, the PAMELA experiment \cite{Adriani:2008zr} measured the positron fraction, $ \phi_{e^+}(E)/(\phi_{e^-}(E)+\phi_{e^+}(E))$,
between $1$ and $100$ GeV, finding that it unexpectedly increases above $10$ GeV. 
This rising behavior is difficult to explain via secondary production of positrons in interactions of  high energy hadronic CRs. 
Therefore it has been interpreted as a positron anomalous excess in the CR energy spectrum above $10$ GeV.  This would imply the existence of an unknown source of CR {positrons}  
- see for instance the nice discussion in \cite{Serpico:2011wg}.  On the other hand PAMELA did not observe any excess in the anti-protons spectrum \cite{Adriani:2010rc}.

Already in 2008 ATIC \cite{:2008zzr} and PPB-BETS \cite{Yoshida:2008zzc} reported an unexpected
structure in $\phi_{e^-}(E)+\phi_{e^+}(E)$, in the energy range between
$100$~GeV and $1$~TeV.
The picture was soon corroborated via the higher-statistics measurements by Fermi-LAT \cite{Abdo:2009zk} and 
H.E.S.S. \cite{Aharonian:2008aa}, that suggested a possible small additional unknown component in the total flux,  
on the top of the standard astrophysical model predictions. The latter generically assume a single-power-law injection 
spectrum of $e^{\pm}$. 
Fermi-LAT \cite{Ackermann:2010ij} 
determined that the total $e^{\pm}$ spectrum in the energy range $7$ GeV$ < E < 1$ TeV
is indeed compatible with a power-law of index $-3.08\pm 0.05$, but it also displays 
significant evidence of a spectral hardening above $100$ GeV.


In 2011, new experimental informations were added.
PAMELA measured the electron spectrum between $1$ and $625$ GeV \cite{Adriani:2011xv}. 
Fermi-LAT \cite{Ackermann:2011rq} measured the separate cosmic-ray electron and positron spectra, thought with a
worse sensitivity than the total spectrum.

Very recently, AMS-02 \cite{AMS02} measured the CR positron fraction with unprecedented precision and up to energies of
about $350$ GeV.
This experimental information, together with the precise determination of the total flux by Fermi-LAT \cite{Ackermann:2010ij}, 
at present, calls for a new study of the charge asymmetry in the electron and positron excesses.
This is what we will perform next following the strategy proposed in \cite{Frandsen:2010mr} using the AMS-02 recent data  
for the CR positron fraction \cite{AMS02}, and Fermi-LAT latest data for the total flux \cite{Ackermann:2010ij}.


\section{Notation and Sum rule}
\label{sumrule}


The observed flux of electrons and positrons can be written as  the sum of two contributions:
a background component $\phi_{e^\pm}^B(E)$, describing all known astrophysical sources; 
an unknown component $\phi_{e^\pm}^U(E)$
(of whatever origin), which is needed to explain the features in the  spectra observed by experiments.
Explicitly,
\beq
\phi_{e^+} (E)= \phi_{e^+}^U (E)+ \phi_{e^+}^B (E) \, ,  \,\,\, \,\, \,\phi_{e^-} (E)= \phi_{e^-}^U (E)+ \phi_{e^-}^B (E) \, .
\eeq
AMS-02 \cite{AMS02} and Fermi-LAT \cite{Ackermann:2010ij} measure respectively the positron fraction and 
the total electron and positron fluxes as a function of the energy $E$: 
\beq
F_+(E) = \frac{\phi_{e^+}(E)}{\phi_{e^+}(E) + \phi_{e^-}(E)}\ , \qquad  T(E) = \phi_{e^+}(E) + \phi_{e^-}(E) \ .
\eeq
The left-hand side of the equations above refer to the experimental measures. 
Given such data,
our aim is to investigate the unknown contribution leading to the lepton excesses:
\begin{eqnarray}
\phi_{e^+}^U(E) & = &  F_+(E)~T(E) -\phi_{e^+}^B(E) \ ,  \\ 
\phi_{e^-}^U(E) & = & T(E)~ \left(1-F_+(E)\right) -\phi_{e^-}^B(E) \ . \nonumber
\end{eqnarray}
Clearly, this can be done only by {\it assuming} an astrophysical background model, 
as discussed below. 

Here we are interested in particular in a fundamental property of the unknown contribution: 
its charge asymmetry \cite{Frandsen:2010mr}. 
The ratio of the unknown electron and positron fluxes is a direct measure of such charge asymmetry: 
\beq
r_U(E) \equiv \frac{\phi_{e^-}^U(E)}{\phi_{e^+}^U(E)}= \frac{T(E) ~(1-F_+(E))-\phi_{e^-}^B(E)}{F_+(E)~T(E)-\phi_{e^+}^B(E)}~~.
\label{ruu}
\eeq
Note that this equation can be rewritten as a {\it sum rule}  \cite{Frandsen:2010mr},
\beq
\frac{T(E)}{  \phi_{e^-}^B(E)  } ~\frac{ 1-(1+r_U(E)) F_+(E)}{1-r_U(E) \frac{\phi_{e^+}^B(E)}{\phi_{e^-}^B(E)}} =1\ ,
\label{sumrulefinal}
\eeq
that links together the experimental results, 
the model of the backgrounds and the dependence on the energy of the charge asymmetry of the unknown excesses.
We use the $E \gtrsim 25$ GeV data bins, since the lower energy bins are affected by the solar modulation. 

\section{Astrophysical Background models}
\label{background}

Primary electrons can come from galactic CRs while interactions of CRs with
the interstellar medium sources secondary electrons, positrons and antiprotons. The propagation of
the signal and background fluxes from their production region to the detector is affected mainly
by diffusion and energy losses. 
We evaluate the background fluxes at Earth using the 
studies \cite{Delahaye:2010ji, Lavalle:2010sf, Lavalle:2012ef, DiBernardo:2012zu}.
These fluxes can be conveniently described by a power low, with a global normalization
and a spectral index. We discuss below the impact of the
  spectral index uncertainties. 

We model the background spectrum using 
\beq
\phi_{e^\pm}^B(E)=N^B_{e^\pm} \,  B_{e^\pm}(E)~~, 
\label{eq:phiB}
\eeq
where $N^{B}_{e^\pm}$ are normalization coefficients 
and $B_{e^\pm}(E)$ are provided using specific astrophysical models. 
In this paper we adopt various models, in order to study how much the results are affected by the background model choice.
In particular, measuring $E$ in ${\rm GeV}$ and the $B$'s in units of ${\rm GeV}^{-1} {\rm cm}^{-2}{\rm sec}^{-1}{\rm sr}^{-1}$,
we consider the following models.

\begin{itemize}

\item Moskalenko and Strong (MS)  \cite{Strong:1998pw}, a popular model 
(used also in \cite{Cirelli:2008id, DeSimone:2013fia})
for which 
\bea
B_{e^+}(E)&=&  \frac{ 4.5 E^{0.7}}{1 + 650 E^{2.3} + 1500 E^{4.2}} \ ,
 \\
B_{e^-}(E)&=& \frac{ 0.16 E^{-1.1}} {1 + 11 E^{0.9} + 3.2 E^{2.15}}  + \frac{ 0.70 E^{0.7}}{1 + 110 E^{1.5} + 600 E^{2.9} + 580 E^{4.2}} \,  .
\nonumber
\eea


\item Spectral Index (SI) model, a generic model that we parametrize as:
\bea
B_{e^+}(E)&=& 1.40 \times 10^{-4}  \left( \frac{E}{E_0} \right)^{-\gamma_{e^+}} \, ,\\
B_{e^-}(E)&=&  5.43 \times 10^{-3} \left( \frac{E}{E_0} \right)^{-\gamma_{e^-}}   \nonumber \, ,
\eea
where $E_0=33.35$ GeV. The normalization coefficients of the SI model have been chosen so that the $B_{e^\pm}$ 
functions equate those of the MS model at $E=E_0$. 
Note also that the SI model with $\gamma_{e^-}=3.21$ and $\gamma_{e^+}=3.41$  actually corresponds to the Fermi Collaboration 
(FC) model zero \cite{Grasso:2009ma, Ibarra:2009dr}. The SI model with $\gamma_{e^+}=3.5$  approximates very well the positron background of the MS model. The electron background of the MS model has spectral index $3.25$ for energies above $100$ GeV.
\end{itemize}

\begin{figure}[h!]
\begin{center} 
\includegraphics[width=12cm]{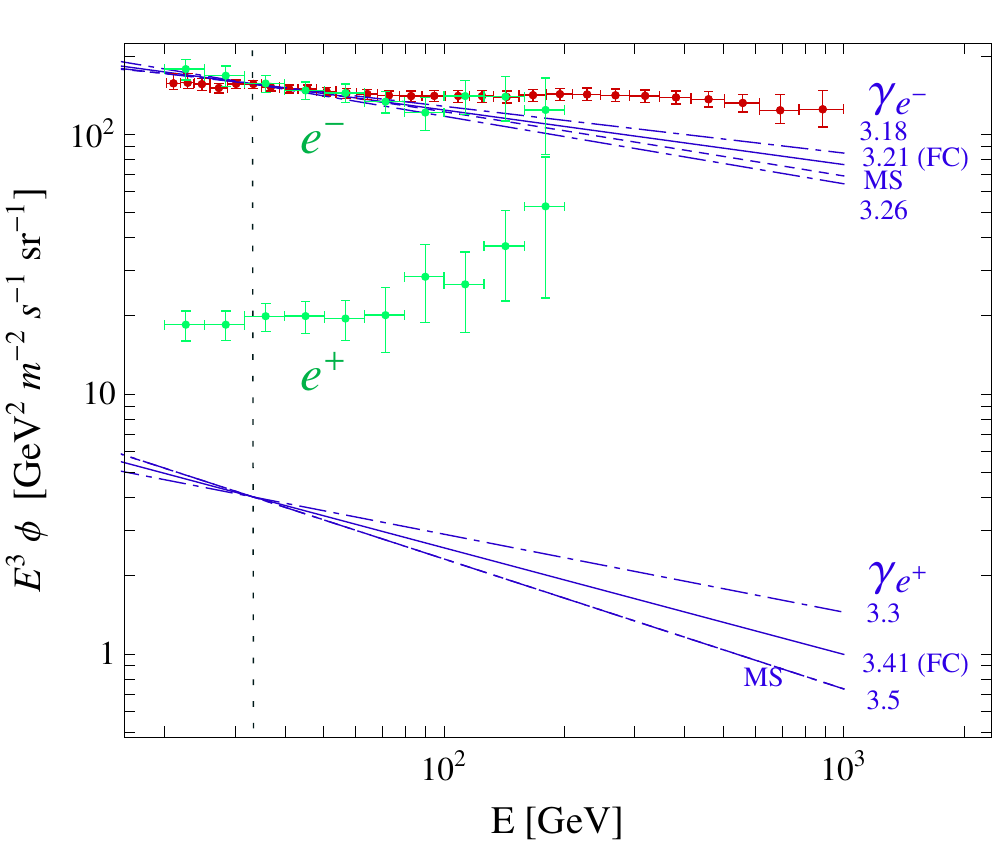}
\end{center}
\vspace*{-0.5cm} 
\caption{\baselineskip=14pt
Models for background fluxes: MS (dashed), SI for extreme (dot-dashed) and intermediate (solid) values of the spectral indexes.
The latter corresponds to the FC model.
The fluxes have been normalized by choosing $N^B_{e^\pm} = 0.73$. 
The Fermi-LAT experimental data on the the total flux \cite{Ackermann:2010ij} (red) and separate electrons and positrons 
fluxes \cite{Ackermann:2011rq} (green) are also shown for comparison.
\baselineskip=16pt}
\label{fig-B}
\end{figure}

The background models are illustrated in fig. \ref{fig-B}, assuming for definiteness $N^B_{e^\pm} = 0.73$. 
The natural range of the electron spectral index of the SI model is $\gamma_{e^-}=[3.18,3.26]$, while the natural 
range of the positron spectral index is $\gamma_{e^+}=[3.4,3.5]$ (see for 
instance \cite{Delahaye:2010ji, Lavalle:2010sf, Lavalle:2012ef}). 
The FC model can thus be seen as an SI model with intermediate values of the spectral indexes.
For comparison, the plot also shows the Fermi-LAT  data points for the total \cite{Ackermann:2010ij}
and separate electron and positron fluxes \cite{Ackermann:2011rq}.
To make a comparison with other recent studies,  we note that refs. \cite{Cirelli:2008id, DeSimone:2013fia} focus 
on the MS model, allowing for variations of its spectral index of about $0.05$; this corresponds to our SI model with
$\gamma_{e^-}=[3.19,3.29]$ and $\gamma_{e^+}=[3.45,3.55]$.

The energy dependence of the background fluxes is encoded in the $B_{e^\pm}(E)$
functions 
but, as can be seen from eq. (\ref{eq:phiB}), there 
is also the problem of fixing the normalization coefficients $N^B_{e^{\pm}}$.
As we are going to discuss, in order to characterize a certain background model, 
 it is not necessary to make two independent assumptions on $N^B_{e^{-}}$ and $N^B_{e^{+}}$, 
 but just one assumption on their ratio: 
 \beq
 r_B = \frac{N^B_{e^-}}{N^B_{e^+}} \,\, .
\eeq
One should not be confused by the use of the variable $r_B$. Note that both primary and secondary
electrons enter the $B_{e^{-}}(E)$ function of eq. (\ref{eq:phiB}), while only secondary positrons contribute
to $B_{e^{+}}(E)$.
So, at the contrary of $r_U(E)$, the ratio $r_B$ has no profound meaning as 
for the issue of charge symmetry: its magnitude expresses just the ratio of the normalization 
coefficients of the assumed astrophysical background fluxes, $N^B_{e^{\pm}}$, which are obtained 
by fitting the low energy data where the excesses are expected to be negligible. Since $N^B_{e^{\pm}}$ 
turn out to be close to unity, the ratio $r_B$ is also expected to be close to unity.   

If we focus on a certain energy bin $\bar E$ where both the positron fraction $F_+$ and the total flux $T$ have 
been measured, eq. (\ref{sumrulefinal}) can indeed be used to derive a range for $N^B_{e^-}$:
\beq
N^B_{e^-}(r_U(\bar E),r_B) =\frac{T(\bar E)}{  B_{e^-}(\bar E)  } ~
\frac{ 1-(1+r_U(\bar E)) F_+(\bar E)}     {1- \frac{ r_U(\bar E)}{r_B}     \frac{B_{e^+}(\bar E)}{B_{e^-}(\bar E) } } \,\, . 
\label{eq:NB}
\eeq
For a certain background model (characterized by the $B_{e^\pm}(E)$ functions) and 
using the experimental data on $T(\bar E)$ and $F_+(\bar E)$, 
the allowed range for $N^B_{e^-}$ can be calculated by making assumptions on the values of $r_U(\bar E)$ and $r_B$. 
For the sake of our analysis, we consider it safe to let $r_B$ vary in the range $[0.5,2]$
(as done also in \cite{Cirelli:2008id, DeSimone:2013fia}).

\begin{figure}[h!]
\vskip 1cm
\begin{center} 
\includegraphics[width=7.6cm]{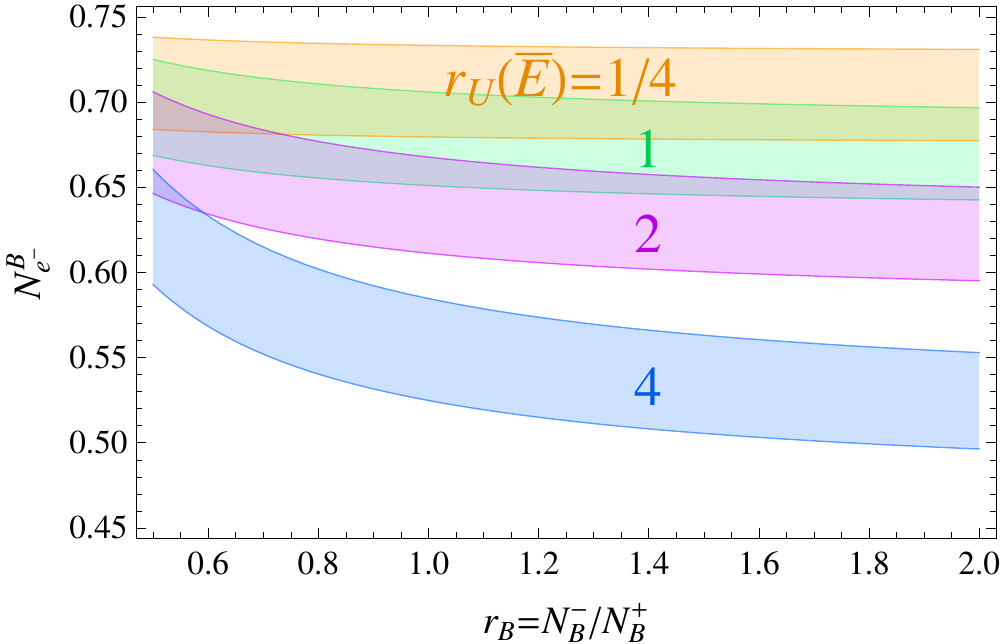}\,\,\,\,\,\quad  \includegraphics[width=7.6cm]{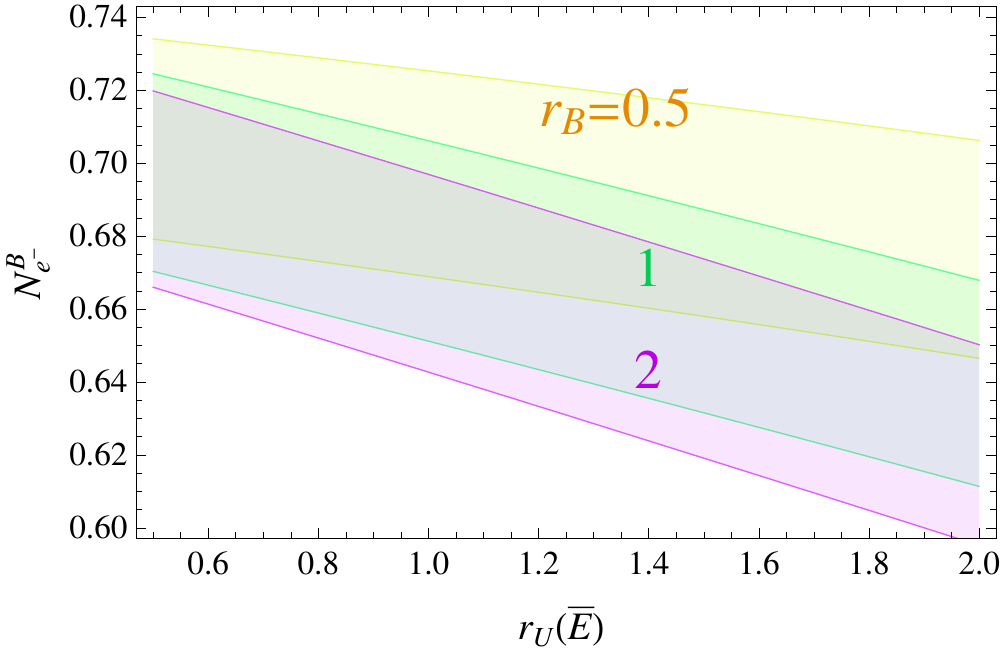} 
\end{center}
\vspace*{-0.5cm} 
\caption{\baselineskip=14pt
Values of $N^B_{e^-}$ according to eq. (\ref{eq:NB}) assuming $\bar E =33.35$ GeV.
Left: as a function of $r_B$ for different values of $r_U(\bar E)$. 
Right: as a function of $r_U(\bar E)$ for different values of $r_B$. 
The thickness of the bands is obtained by considering the $1\sigma$ ranges  
of $F_+(\bar E)$ from AMS-02 \cite{AMS02} and $T(\bar E)$ from Fermi-LAT \cite{Ackermann:2010ij}. 
\baselineskip=16pt}
\label{fig-NB1}
\end{figure}

We consider in particular $\bar E=33.35$ GeV (for which $B_{e^-}(\bar E)/B_{e^+}(\bar E)= 38.82$)
and display the results for $N^B_{e^-}$ in fig. \ref{fig-NB1}, showing the dependence on $r_B$ 
for fixed values of $r_U(\bar E)$ in the left panel, viceversa in the right panel.
The thickness of the curves is obtained by considering the variation of $N^B_{e^-}$
associated to the $1\sigma$ ranges of $T(\bar E)$ and $F_+(\bar E)$. 
The variation due to $T(\bar E)$ turns out to be the dominant one.

%

\section{Constraining the Charge asymmetry}
\label{asymm}

Having derived a range for the normalization of the electron background $N^B_{e^-}$ as a function of $r_U(\bar E)$ and $r_B$
(so that both $T$ and $F_+$ values are reproduced at a certain energy $\bar E$), 
we can use it to extrapolate the positron fraction at any energy value.
We can in fact rewrite eq. (\ref{sumrulefinal}) as follows:
\begin{equation}
F_+(E) = \frac{1}{1+r_U(E)} \left[1-N^B_{e^-}(r_U(\bar E),r_B)\, \frac{B_{e^-}(E)}{T(E)} \left(1-\frac{r_U(E)}{r_B} 
\frac{B_{e^+}(E)}{B_{e^-}(E) }  \right)\right] \, .
\label{extrap}
\end{equation}
Suppose now that we specify a background model (namely $B_{e^+}(E),B_{e^-}(E),r_B$) and that we 
make an assumption on $r_U(E)$: it is then
possible to check the consistency between the extrapolation for different energy values based on eq. (\ref{extrap}) 
and the experimental data. 
Clearly, if we use the range of values for $N^B_{e^-}(r_U(\bar E),r_B)$ discussed previously, 
we are guaranteed that both the total flux and the positron fraction reproduce the experimental data in the $\bar E$ energy bin.

\subsection{Energy independent $r_U(E)$}

In order to test whether current CRs data could support charge asymmetric lepton excesses, 
as done in \cite{Frandsen:2010mr},
the first step is to consider the oversimplifying assumption that $r_U$ is nearly constant in the energy region of interest.

\begin{figure}[h!]
\begin{center} 
\includegraphics[width=11cm]{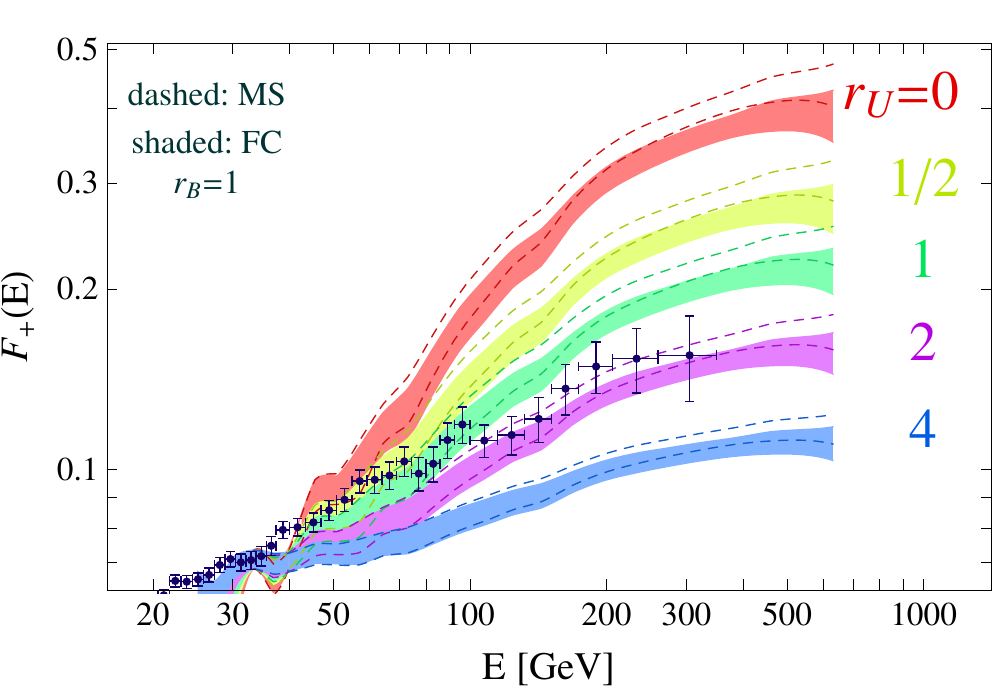} \end{center}
\vspace*{-0.5cm} 
\caption{\baselineskip=14pt
Positron fraction $F_+(E)$ for different values of $r_U$, according to eq. (\ref{extrap}). 
The shaded (dashed) curves refer to the FC (MS) background model with $r_B=1$.
The AMS-02 data \cite{AMS02} with 1$\sigma$ error bars (statistical and systematic combined in quadrature) 
are shown for comparison.\baselineskip=16pt}
\label{predictionPF}
\end{figure}

The extrapolation of the positron fraction,  assuming that $r_U$ remains constant over the entire energy range 
(from about $30$ GeV up to about $700$ GeV), 
is shown in fig.~\ref{predictionPF} for the FC (shaded) and MS (dashed) background models and by taking $r_B=1$. 
We focus in particular on the cases $r_U=0,1/2,1,2,4$ (from top to bottom).
The thickness of the curves correspond to the $1\sigma$ ranges of AMS-02 \cite{AMS02} and Fermi-LAT \cite{Ackermann:2010ij}.
One should not  be worried by the wiggles between in the $30-50$ GeV, as they are simply due to the features of the data points 
in that range. 
The picture shows that the current data are consistent with $r_U =1$ for energies up to $100$ GeV, but
favor a deviation from charge symmetry  at energies above $100$ GeV, where the preferred charge asymmetry value is $r_U \sim 2$:
the latter value means that the  electron excess of unknown origin should be about twice the positron one.

These interesting results makes it mandatory a deeper study of their dependence on the model background:
this can be done by considering the impact of varying $r_B$ and the spectral indexes $\gamma_{e^\pm}$. 

\begin{figure}[h!]
\begin{center} 
\includegraphics[width=9cm]{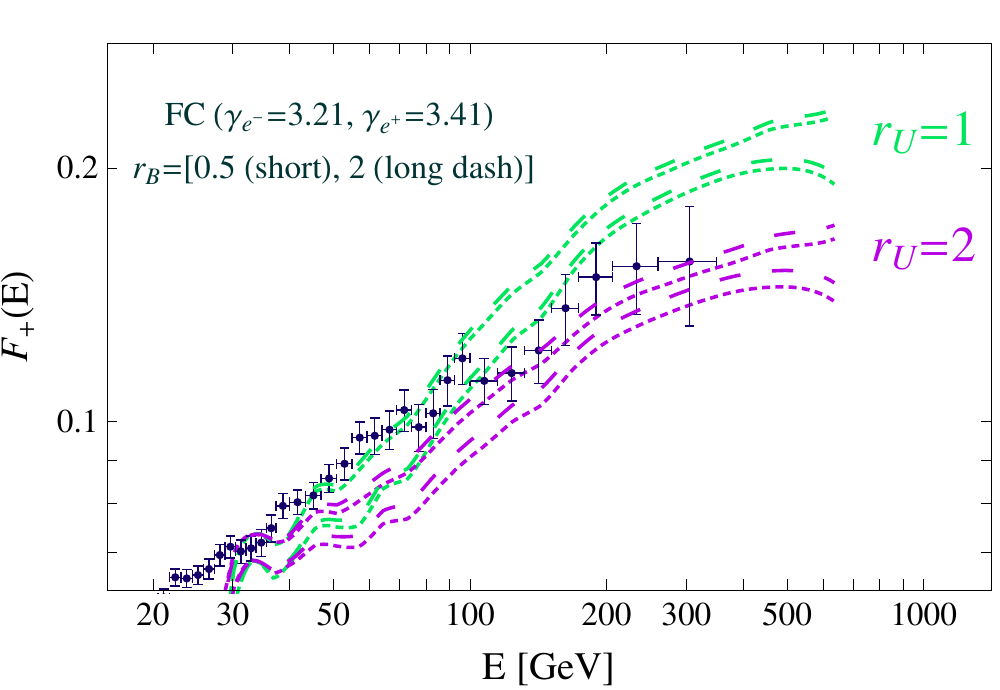}
\includegraphics[width=9cm]{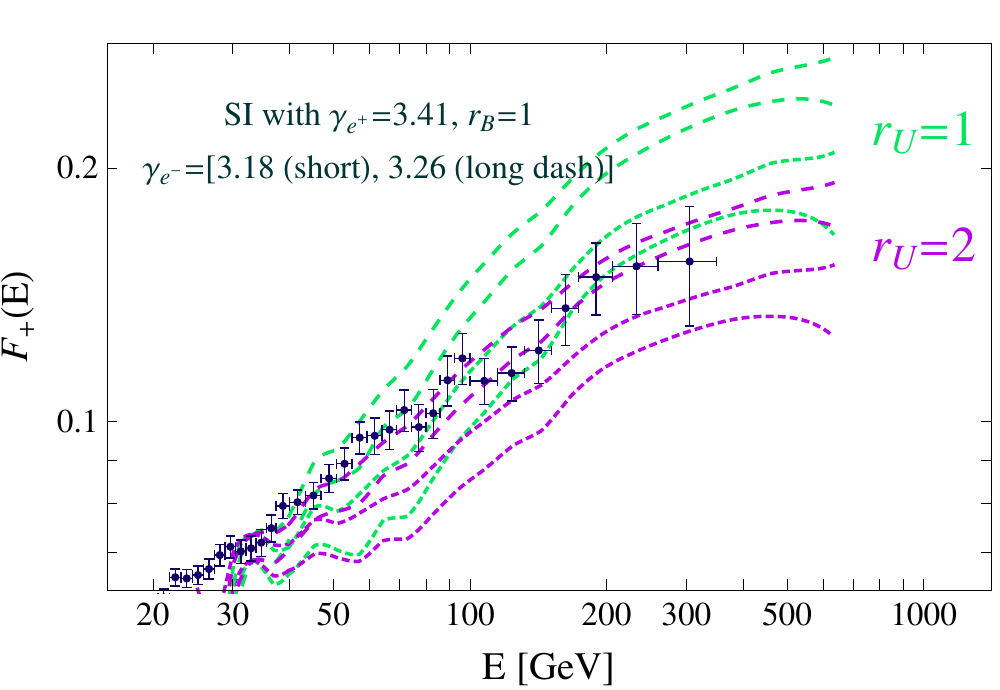}   
\includegraphics[width=9cm]{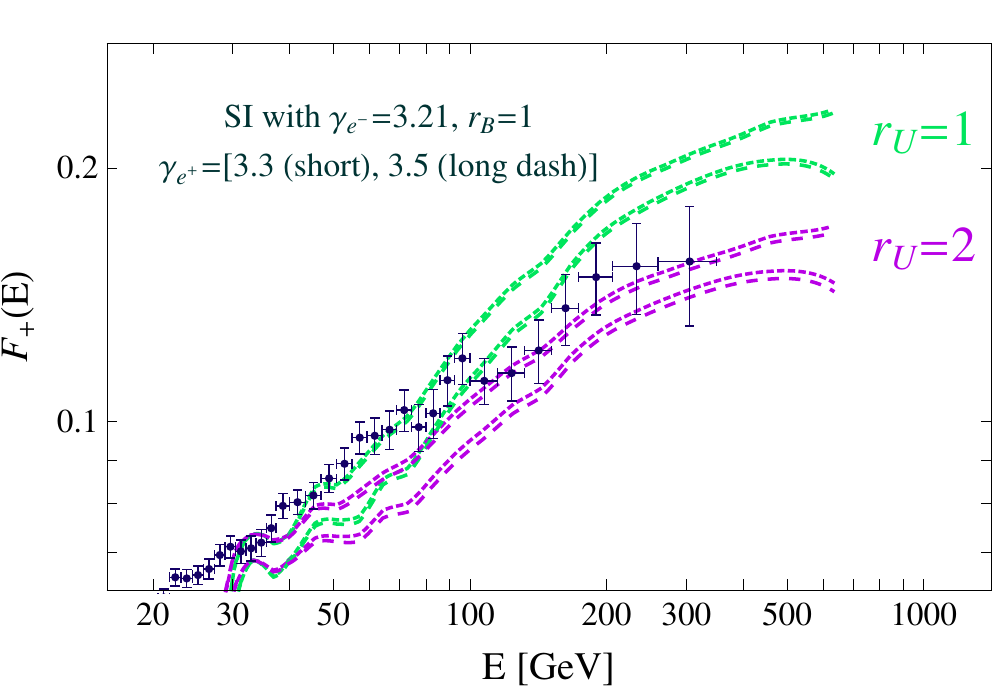} 
\end{center}
\vspace*{-0.5cm} 
\caption{\baselineskip=14pt
Positron fraction $F_+(E)$ for different values of $r_U$. 
The AMS-02 data \cite{AMS02} with 1$\sigma$ error bars  are shown for comparison.
Top:  dependence on $r_B$ for FC model. 
Medium: dependence on $\gamma_{e^-}$ for SI model with $\gamma_{e^+}=3.41$ and $r_B=1$. 
Bottom: dependence on $\gamma_{e^+}$ for SI model with $\gamma_{e^-}=3.21$ and $r_B=1$. 
\baselineskip=16pt}
\label{predictionPFbis}
\end{figure}

Focusing on the FC background model for definiteness, the top panel of fig. \ref{predictionPFbis} displays the dependence on $r_B$. 
Lowering $r_B$ goes in the direction of reducing the positron fraction, alleviating the tension between the AMS-02 data \cite{AMS02} 
and the charge symmetric case above $100$ GeV. However, the global shift turns out not to be strong enough to account for $r_U=1$. 

As a second study of the robustness of the deviation from charge symmetry above $100$ GeV, 
we consider a generic SI model with  $r_B=1$ and analyze its dependence on $\gamma_{e^-}$ and $\gamma_{e^+}$:
this is done respectively in the middle and bottom panels of fig.  \ref{predictionPFbis}. 
We can see that while $\gamma_{e^-}$ has a significative impact on the slope of the positron fraction,
$\gamma_{e^+}$ does not affect it much. 
The tension with charge symmetry above $100$ GeV is nearly removed for values of  $\gamma_{e^-}$ smaller than $3.18$.

As a further test of our results, we perform a chi-squared test using the following test function. 
For the background we assume the following SI form, 
\beq
\phi_{e^\pm}^B(E)=N^B_{e^\pm} \,  E^{-\gamma_{e^\pm}}\, ,
\eeq
while for the unknown source we consider
\beq
\phi_{e^\pm}^U(E)=N^U_{e^\pm} \,  E^{-\gamma_{U}} \, e^{-E/E_U}\,\,.
\label{eq:fit}
\eeq
The associated  charge asymmetry is constant and parametrized by $r_U=N^U_{e^-}/N^U_{e^+}$.
We fit the AMS-02 \cite{AMS02} and Fermi-LAT data \cite{Ackermann:2010ij}
to the following seven free parameters for a given $r_U$: 
\beq
\frac{N^B_{e^-}}{N^B_{e+}},\,\,  \gamma_{e^-}-\gamma_{e^+},\, \, \frac{N^U_{e^-}}{N^B_{e^-}}, \,\,
\gamma_{U}- \gamma_{e^-},\,  \,E_U, \, \, N^B_{e^-}, \, \,\gamma_{e^-} \, \, , \quad {\rm free \, parameters} \, .
\eeq
A reasonable agreement with data, according to the chi-squared distribution, can be obtained for any 
value of  $r_U$ between $1$ and $2$. 
We find that for lower values of $r_U$, $\gamma_{e^-}$ turns out to be small, around $3.14-3.16$ at $90\%$  C.L. for $r_U=1$.
For $r_U=2$ instead the $\gamma_{e^-}$ range is $3.205-3.225$.
These results are in agreement with our previous comments that a  value of 
$\gamma_{e^-}$ closer to the astrophysically expected range $3.18-3.26$, is obtained for $r_U \sim 2$.
Turning the argument around, assuming the FC model with $\gamma_{e^-}=3.21$ and $\gamma_{e^+}=3.41$, 
we obtain that the combined ranges for $r_U$ and $E_U$ are respectively
$1.7-1.9$ and $600-1500$ GeV at $90\%$ C.L..
 
\subsection{Deriving $r_U(E)$ from data}

For various background models, 
we now study directly the charge asymmetry of the unknown excesses, 
$r_U(E)= \phi_-^U(E)/\phi_+^U(E)$, by considering its expression given in eq. (\ref{ruu}), which we report here:
\beq
r_U(E) = \frac{T(E) 
~(1-F_+(E))-  N^{B}_{e^-}(r_U(\bar E),r_B) \,B_{e^-}(E)}{F_+(E)~T(E)-\frac{N^{B}_{e^-}(r_U(\bar E),r_B)}{r_B} \,B_{e^+}(E)}~~.
\label{eq:rUE}
\eeq
Such study can be done by assuming a background model (namely $B_{e^+}(E),B_{e^-}(E),r_B$) and making an
assumption on $r_U(\bar E)$.

\begin{figure}[h!]
\begin{center} \vskip .1cm
\includegraphics[width=10.cm]{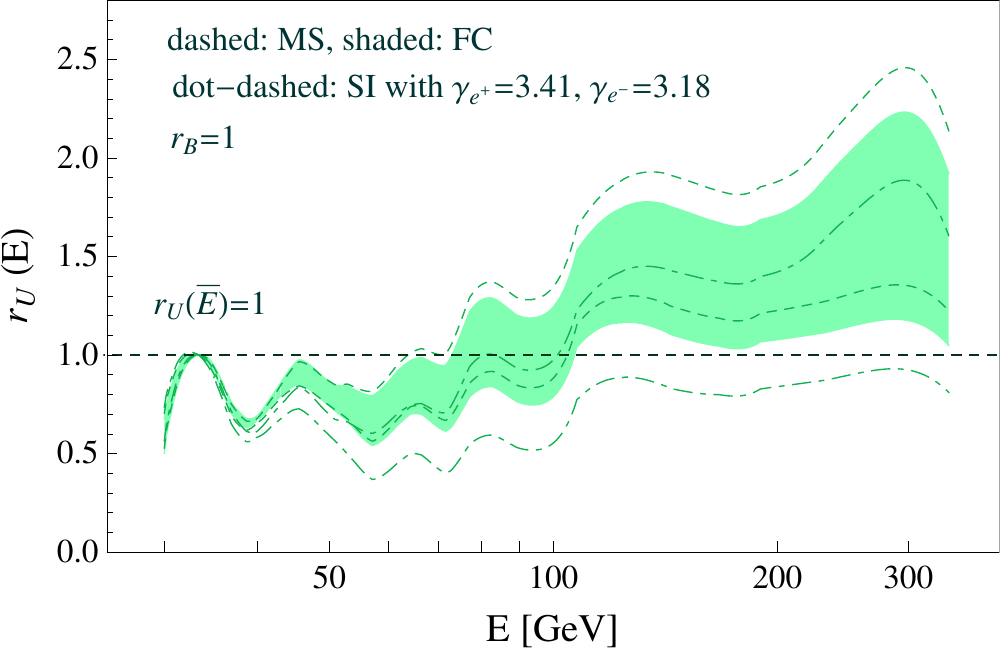} \vskip 1cm
\includegraphics[width=10.cm]{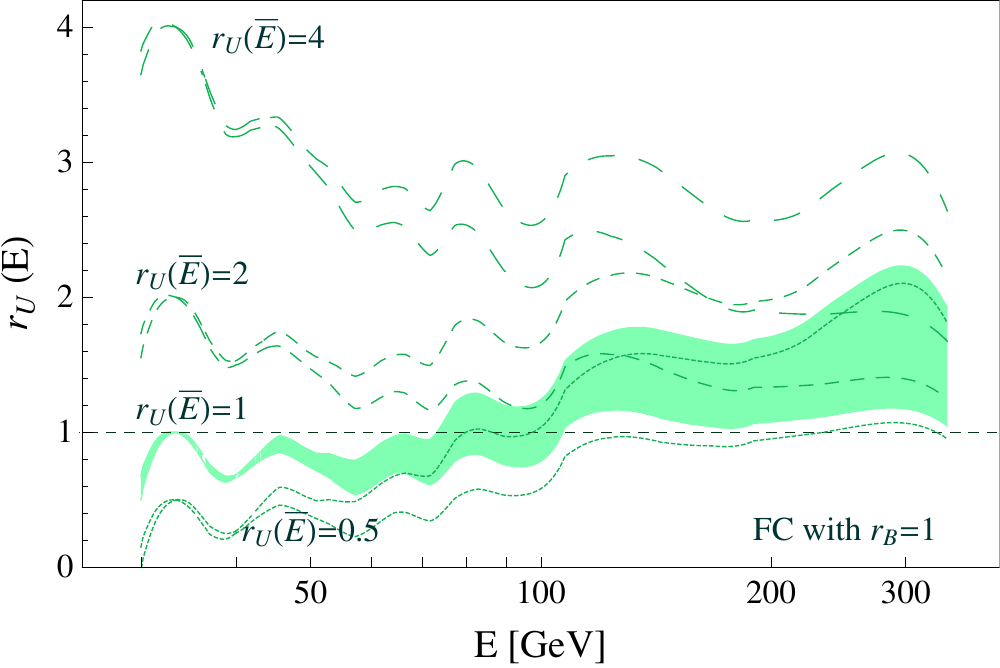}  \end{center}
\vspace*{-0.5cm} 
\caption{\baselineskip=14pt
Ratio $r_U(E)$ according to eq. (\ref{eq:rUE}). 
Top: for the MS (dashed), FC (shaded) and SI (dot-dashed) models with $r_B=1$ and $r_U(\bar E)=1$. 
Bottom: for the FC model with $r_B=1$ and various values of $r_U(\bar E)$.
The thickness of the curves corresponds to the $1\sigma$ ranges of AMS-02 \cite{AMS02} and Fermi-LAT \cite{Ackermann:2010ij}.
\baselineskip=16pt}
\label{fig-ru}
\end{figure}

In the top panel of fig. \ref{fig-ru}  we display the $r_U(E)$ range for the MS (dashed), FC (shaded) and SI (dot-dashed) models, assuming $r_B=1$ and $r_U(\bar E)=1$. For the SI model we choose $\gamma_{e^-}=3.18$ and $\gamma_{e^+}=3.41$.
The thickness of the curves corresponds to the $1\sigma$ ranges of AMS-02 \cite{AMS02} and Fermi-LAT \cite{Ackermann:2010ij}.
Despite the oscillations below $50$ GeV (which are simply due to the pattern of the experimental data points), 
one can see that $r_U(E)$ displays an increasing behavior with energy. A transition occurs above $100$ GeV, where
$r_U(E)$ becomes bigger than unity, spanning the range between $1$ and $2$, for the MS and FC models. The SI model with a low
value of the electron spectral index, $\gamma_{e^-}=3.18$, is instead compatible with charge symmetry.

Since we are particularly interested in deviations from charge symmetry, it is important to study the dependence of $r_U(E)$
on the value chosen for $r_U(\bar E)$. This is done in the bottom panel of fig. \ref{fig-ru}, 
focusing on the FC model with $r_B=1$.
One can see that for whole interval $r_U(\bar E)=[0.5,4]$, $r_U(E)$ is bigger the unity above $100$ GeV. In addition,
$r_U(\bar E)\sim 2$ is the sole case that allows $r_U(E)$ to be nearly energy independent. 
For $r_U(\bar E)> 2$ ($r_U(\bar E)< 2$), $r_U(E)$ is a  decreasing (increasing) function of the energy.
We also considered the dependence of the $r_U(E)$ on the value chosen for $r_B$, finding it to be negligible.

%
%
%
%
%

\section{Applications and Conclusions}
\label{conclusions}
The physical  sources of the excesses observed in the electron and positron CR fluxes are naturally divided in two classes:
charge symmetric and charge asymmetric. 
The first class is characterized by $r_U = 1$. All other sources, for which $r_U(E)$ depends on the energy or is constant
but different from unity, belong to the second class.

This  classification applies indiscriminately to any source, be it astrophysical or of DM nature.
It is not the goal of this paper to enter in the merit of any specific model. Here we merely classify the main 
models suggested in the literature with respect to their potential to yield a charge asymmetry.
Then we suggest how to use our results about charge asymmetry, expressed in fig. \ref{fig-ru},
for a straightforward test of any production mechanism.

\vskip .3cm
{\bf Astrophysical models}\\
The simplest models of pulsars -- see {\it e.g.}  \cite{Linden:2013mqa} and references therein --
are charge symmetric, given that the basic assumption is that pulsars inject the same number of electrons and 
positrons in the interstellar medium. 
Our parameterization eq. (\ref{eq:fit}) with $N^U_{e^+}=N^U_{e^-}$ (hence $r_U=1$)
describes well the  fluxes from a pulsar expected at Earth.

For supernovae \cite{Delahaye:2010ji, Serpico:2011wg, Panov:2013fma,Caprioli:2009fv} the situation is more delicate. 
There is a primary source of electrons which is responsible for $B_{e^-}(E)$ 
and, on top of that, an equal number of positron and electron secondaries produced at the source 
and which might be responsible for the excesses. Here too the production mechanism for the excesses should
correspond to $r_U=1$.

\vskip .3cm
{\bf DM models} \\
Symmetric DM can lead to CRs by either annihilation, decay or both. 
However, as proven in \cite{Masina:2011hu}, all models of symmetric DM imply $r_U=1$ for any energy. 
A recent re-analysis has been performed in ref.  \cite{DeSimone:2013fia}.

In order to achieve an  $r_U\neq 1$, DM must be asymmetric 
(therefore decaying) \cite{Nussinov:1985xr, Barr:1990ca, Gudnason:2006yj,Nardi:2008ix,Kaplan:2009ag}
and furthermore violate lepton flavor symmetry \cite{Masina:2011hu, Chang:2011xn, Carone:2011ur}.  
These models  lead to an energy dependent $r_U$. 
For instance, for an asymmetric DM candidate decaying into $\mu^- \tau^+$ 
we obtain a naturally increasing behavior for $r_U(E)$, from about $1.5$ at $E=30$ GeV up to a value of $3$  at $E=300$ GeV,
see fig. 3 of \cite{Masina:2011hu}.

For the DM interpretation of the excesses, however, attention must be paid to gamma-ray constraints, see for instance \cite{Cirelli:2010xx}.

\vskip .5cm

We have shown that by combining the recent data from AMS-02 with those from Fermi-LAT,
a charge asymmetry  in the unknown excesses of electron and positron CRs is emerging,
favoring the electron component.
The result relies on having chosen a conservative estimate for the astrophysical background fluxes.
Charge symmetry can be rescued when adopting an electron background spectral index
slightly smaller than the commonly assumed values. 

Given that both astrophysical and DM models can be classified according to their predictions for the asymmetry, 
the impact of the future AMS-02 \cite{AMS02} 
results for the charge asymmetry will play a crucial role in discriminating the proposed production mechanisms.

\acknowledgements
I. M. thanks P. Blasi for useful discussion.
The CP$^3$-Origins centre is partially founded by the Danish National Research Foundation under the research grant DNRF:90.



\begin{thebibliography}{99}

 
\bibitem{Fan:2010yq}
  Y.~Z.~Fan, B.~Zhang and J.~Chang,
  Int.\ J.\ Mod.\ Phys.\  {\bf D19 } (2010)  2011-2058.
  
\bibitem{Panov:2013fma}
  A.~D.~Panov,
  Journal of Physics: Conference Series {\bf 409} (2013) 012004
  [arXiv:1303.6118 [astro-ph.HE]].


\bibitem{Frandsen:2010mr}
  M.~T.~Frandsen, I.~Masina, F.~Sannino,
    Phys.\ Rev.\  D {\bf 83} (2011) 127301
  [arXiv:1011.0013 [hep-ph]].
 See also I.~Masina,
  [arXiv:1105.0089 [hep-ph]]. 
  

\bibitem{AMS02} 
S. Ting, ÒRecent results from the AMS experimentÓ, seminar given at CERN on 3rd April 2013; 
M.Aguilar et al. (AMS Collaboration), Phys. Rev. Lett. 110, 141102 (2013).
Main results available from http://www.ams02.org/. 

\bibitem{Ackermann:2010ij}
  M.~Ackermann {\it et al.}  [Fermi LAT Collaboration],
  Phys.\ Rev.\  D {\bf 82} (2010) 092004
  [arXiv:1008.3999 [astro-ph.HE]].
  
\bibitem{Ackermann:2011rq}
 M. Ackermann {\it et al.} [ The Fermi LAT Collaboration ],
  [arXiv:1109.0521 [astro-ph.HE]].  


\bibitem{Delahaye:2010ji}
  T.~Delahaye, J.~Lavalle, R.~Lineros, F.~Donato and N.~Fornengo,
  Astron.\ Astrophys.\  {\bf 524} (2010) A51
  [arXiv:1002.1910 [astro-ph.HE]].
  
  
\bibitem{Serpico:2011wg}
  P.~D.~Serpico,
  Astropart.\ Phys.\  {\bf 39-40} (2012) 2
  [arXiv:1108.4827 [astro-ph.HE]].
  
  
  
\bibitem{Linden:2013mqa}
  T.~Linden and S.~Profumo,
  arXiv:1304.1791 [astro-ph.HE].

\bibitem{Caprioli:2009fv}
  D.~Caprioli, E.~Amato and P.~Blasi,
  Astropart.\ Phys.\  {\bf 33} (2010) 160
  [arXiv:0912.2964 [astro-ph.HE]];
  P.~Blasi,
  Phys.\ Rev.\ Lett.\  {\bf 103} (2009) 051104
  [arXiv:0903.2794 [astro-ph.HE]].



\bibitem{Masina:2011hu}
  I.~Masina and F.~Sannino,
  JCAP {\bf 1109} (2011) 021
  [arXiv:1106.3353 [hep-ph]].

\bibitem{Chang:2011xn}
  S.~Chang and L.~Goodenough,
  Phys.\ Rev.\ D {\bf 84} (2011) 023524
  [arXiv:1105.3976 [hep-ph]].
  

\bibitem{Carone:2011ur}
  C.~D.~Carone, A.~Cukierman and R.~Primulando,
  Phys.\ Lett.\ B {\bf 704} (2011) 541
  [arXiv:1108.2084 [hep-ph]].

%
  
 
\bibitem{Adriani:2008zr}
  O.~Adriani {\it et al.}  [PAMELA Collaboration],
  Nature {\bf 458} (2009) 607
  [arXiv:0810.4995 [astro-ph]].
  O.~Adriani {\it et al.} [PAMELA Collaboration],
  Astropart.\ Phys.\  {\bf 34} (2010) 1
  [arXiv:1001.3522 [astro-ph.HE]].




\bibitem{Adriani:2010rc}
  O.~Adriani {\it et al.}  [PAMELA Collaboration],
  Phys.\ Rev.\ Lett.\  {\bf 105} (2010) 121101
  [arXiv:1007.0821 [astro-ph.HE]].
See also:  O.~Adriani {\it et al.},
  Phys.\ Rev.\ Lett.\  {\bf 102} (2009) 051101
  [arXiv:0810.4994 [astro-ph]].
  
 
  
\bibitem{:2008zzr}
  J.~Chang {\it et al.} [ATIC Collaboration],
  Nature {\bf 456} (2008) 362.
 
\bibitem{Yoshida:2008zzc}
  K.~Yoshida {\it et al.},
  Adv.\ Space Res.\  {\bf 42}, 1670 (2008).
  S.~Torii {\it et al.}  [PPB-BETS Collaboration],
  arXiv:0809.0760 [astro-ph]. 
  
  
\bibitem{Abdo:2009zk}
  A.~A.~Abdo {\it et al.}  [Fermi LAT Collaboration],
  Phys.\ Rev.\ Lett.\  {\bf 102} (2009) 181101
  [arXiv:0905.0025 [astro-ph.HE]].


  
\bibitem{Aharonian:2008aa}
  F.~Aharonian {\it et al.}  [H.E.S.S. Collaboration],
  Phys.\ Rev.\ Lett.\  {\bf 101} (2008) 261104.
  F.~Aharonian {\it et al.} [H.E.S.S. Collaboration],
  Astron.\ Astrophys.\  {\bf 508 } (2009)  561
[arXiv:0905.0105]. 
    

  

  
\bibitem{Adriani:2011xv}
  O.~Adriani {\it et al.} [ PAMELA Collaboration ],
  Phys.\ Rev.\ Lett.\  {\bf 106}, 201101 (2011).
  [arXiv:1103.2880 [astro-ph.HE]]. 
  
  

 
\bibitem{Lavalle:2010sf}
  J.~Lavalle,
  Mon.\ Not.\ Roy.\ Astron.\ Soc.\  {\bf 414} (2011) 985L
  [arXiv:1011.3063 [astro-ph.HE]].
  
\bibitem{Lavalle:2012ef}
  J.~Lavalle and P.~Salati,
  Comptes Rendus Physique {\bf 13} (2012) 740
  [arXiv:1205.1004 [astro-ph.HE]].
  
\bibitem{DiBernardo:2012zu}
  G.~Di Bernardo, C.~Evoli, D.~Gaggero, D.~Grasso and L.~Maccione,
  JCAP {\bf 1303} (2013) 036
  [arXiv:1210.4546 [astro-ph.HE]].


\bibitem{Strong:1998pw}
  A.~W.~Strong and I.~V.~Moskalenko,
  Astrophys.\ J.\  {\bf 509} (1998) 212.
  E.~A.~Baltz and J.~Edsjo,
  Phys.\ Rev.\  D {\bf 59} (1998) 023511.
  

\bibitem{Cirelli:2008id}
  M.~Cirelli, R.~Franceschini and A.~Strumia,
  Nucl.\ Phys.\ B {\bf 800} (2008) 204
  [arXiv:0802.3378 [hep-ph]].
  M.~Cirelli, M.~Kadastik, M.~Raidal and A.~Strumia,
  Nucl.\ Phys.\ B {\bf 813} (2009) 1
  [arXiv:0809.2409 [hep-ph]].
  
\bibitem{DeSimone:2013fia}
  A.~De Simone, A.~Riotto and W.~Xue,
  arXiv:1304.1336 [hep-ph].

  
%

\bibitem{Grasso:2009ma}
  D.~Grasso {\it et al.}  [FERMI-LAT Collaboration],
  Astropart.\ Phys.\  {\bf 32} (2009) 140.

\bibitem{Ibarra:2009dr}
  A.~Ibarra, D.~Tran and C.~Weniger,
  JCAP {\bf 1001} (2010) 009.
  
  
\bibitem{Nussinov:1985xr} 
  S.~Nussinov,
  Phys.\ Lett.\ B {\bf 165}, 55 (1985).
  
\bibitem{Barr:1990ca} 
  S.~M.~Barr, R.~S.~Chivukula and E.~Farhi,
  Phys.\ Lett.\ B {\bf 241}, 387 (1990).

\bibitem{Gudnason:2006yj}
  S.~B.~Gudnason, C.~Kouvaris and F.~Sannino,
  Phys.\ Rev.\ D {\bf 74} (2006) 095008
  [hep-ph/0608055].

\bibitem{Nardi:2008ix}
  E.~Nardi, F.~Sannino and A.~Strumia,
  JCAP {\bf 0901} (2009) 043
  [arXiv:0811.4153 [hep-ph]].
 
\bibitem{Kaplan:2009ag}
  D.~E.~Kaplan, M.~A.~Luty and K.~M.~Zurek,
  Phys.\ Rev.\ D {\bf 79} (2009) 115016
  [arXiv:0901.4117 [hep-ph]].
  

%
%
%
%
%
%
  
\bibitem{Cirelli:2010xx}
  M.~Cirelli, G.~Corcella, A.~Hektor, G.~Hutsi, M.~Kadastik, P.~Panci, M.~Raidal and F.~Sala {\it et al.},
  JCAP {\bf 1103} (2011) 051
   [Erratum-ibid.\  {\bf 1210} (2012) E01]
  [arXiv:1012.4515 [hep-ph]].
M.~Cirelli, E.~Moulin, P.~Panci, P.~D.~Serpico and A.~Viana,
  Phys.\ Rev.\ D {\bf 86} (2012) 083506
  [arXiv:1205.5283 [astro-ph.CO]].
I.~Masina, P.~Panci and F.~Sannino,
  JCAP {\bf 1212} (2012) 002
  [arXiv:1205.5918 [astro-ph.CO]].  

 \end{thebibliography}
 \end{document}